\documentclass[]{pasj01}

\begin{document} 
\Received{2016/12/30}
\Accepted{2017/04/08}

\title{The shortest periodic and flaring flux variability 
of a methanol maser emission at 6.7~GHz in G~014.23$-$00.50} 

\author{Koichiro \textsc{Sugiyama}\altaffilmark{1,2$\ast$}}
\altaffiltext{1}{Center for Astronomy, Ibaraki University, 
2-1-1 Bunkyo, Mito, Ibaraki 310-8512, Japan}
\email{koichiro.sugiyama@nao.ac.jp}

\author{Katsura \textsc{Nagase}\altaffilmark{3}}

\author{Yoshinori \textsc{Yonekura}\altaffilmark{1}}

\author{Munetake \textsc{Momose}\altaffilmark{1,3}}

\author{Yasutaka \textsc{Yasui}\altaffilmark{3}}

\author{Yu \textsc{Saito}\altaffilmark{1}}

\author{Kazuhito \textsc{Motogi}\altaffilmark{4}}

\author{Mareki \textsc{Honma}\altaffilmark{5}}

\author{Kazuya \textsc{Hachisuka}\altaffilmark{5}}

\author{Naoko \textsc{Matsumoto}\altaffilmark{2,6}}

\author{Mizuho \textsc{Uchiyama}\altaffilmark{7}}

\author{Kenta \textsc{Fujisawa}\altaffilmark{6}}

\altaffiltext{2}{Mizusawa VLBI Observatory, National Astronomical 
Observatory of Japan (NAOJ), 2-21-1 Osawa, Mitaka, Tokyo 181-8588, Japan}

\altaffiltext{3}{College of Science, Ibaraki University, 
2-1-1 Bunkyo, Mito, Ibaraki 310-8512, Japan}

\altaffiltext{4}{Graduate School of Sciences and Technology for Innovation, 
Yamaguchi University, 1677-1 Yoshida, Yamaguchi, 
Yamaguchi 753-8512, Japan}

\altaffiltext{5}{Mizusawa VLBI Observatory, NAOJ, 2-12 Hoshigaoka-cho, 
Mizusawa-ku, Oshu, Iwate 023-0861, Japan}

\altaffiltext{6}{The Research Institute for Time Studies, 
Yamaguchi University, 1677-1 Yoshida, Yamaguchi, 
Yamaguchi 753-8511, Japan}

\altaffiltext{7}{Advanced Technology Center, NAOJ, 2-21-1 Osawa, 
Mitaka, Tokyo 181-8588, Japan}


\maketitle

\begin{abstract}
We detected flaring flux variability that regularly occurred with 
the period of 23.9~days on a 6.7~GHz methanol maser emission 
at $V_{\mathrm{lsr}} =$ 25.30~km~s$^{-1}$ in G~014.23$-$00.50 
through highly frequent monitoring using the Hitachi 32-m radio telescope. 
By analyzing data 
from 05 January 2013 to 21 January 2016, 
the periodic variability has persisted 
in at least 47 cycles, corresponding to $\sim$1,100~days. 
The period of 23.9~days is the shortest one observed in masers 
at around high-mass young stellar objects so far. 
The flaring component normally falls below 
the detection limit (3$\sigma$) of $\sim$0.9~Jy. 
In the flaring periods, 
the component rises above the detection limit 
with the ratio of the peak flux density more 
than 180 in comparison with a quiescent phase, 
showing intermittent periodic variability. 
The time-scale of the flux rise was typically two days or shorter, and 
both symmetric and asymmetric profiles of flux variability were 
observed through intraday monitoring. 
These characteristics might be explained by 
a change in the flux of seed photons by a colliding-wind binary (CWB) system 
or a variation of the dust temperature by an extra heating source of a shock 
formed by the CWB system within a gap region in a circumbinary disk, 
in which the orbital semi-major axes of the binary are 0.26--0.34~au. 
\end{abstract}

\section{Introduction}\label{section1}

Periodic flux variability of the methanol masers was first discovered 
in G~009.62$+$00.19~E, 
i.e., periodic variability with the interval of 246 days (\cite{goedhart03}; 
modified to $243.3 \pm2.1$~days in \cite{goedhart14}). 
So far, the periodic flux variability of the methanol masers has been 
detected in 20 sources (including quasi-periodic ones). 
Their periods range from 29.5 to 668 days 
(\cite{goedhart04}, \yearcite{goedhart09}; \cite{araya10}; 
\cite{szymczak11}, \yearcite{szymczak14}, \yearcite{szymczak15}, \yearcite{szymczak16}; 
\cite{fujisawa14a}; \cite{maswan15}, \yearcite{maswan16}). 
Patterns of the variability have been classified into 
two categories: the sinusoidal one, and the intermittent one with a quiescent phase. 
Such periodic flux variability was also observed in other masers, 
i.e., 
silicon monoxide in Orion Kleinmann-Low (KL) \citep{ukita81}, 
formaldehyde in IRAS~18556$+$0408 \citep{araya10}, 
hydroxyl in G~012.88$+$00.48 \citep{green12a}, 
and water in IRAS~22198$+$6336 \citep{szymczak16}, 
and variations of these lines except for the silicon monoxide 
were synchronized with the variations of methanol masers 
in the same sources. 
The periodic variability, therefore, must be a common phenomenon 
at around high-mass (proto-)stars, but appears in limited conditions. 
Because of their short timescale, the periodic variability 
is potentially important in studying high-mass protostars 
and their circumstellar structure on spatial scales of 0.1--1~au, 
which are estimated under the condition of Keplerian rotation. 

Four models have been proposed for interpretations of the periodic flux 
variability, in the point of view that 
variations of (multiple) spectral components are synchronized 
with 1--14 days' delay in some sources, possibly caused by global 
variation on a central engine: 
a colliding-wind binary (CWB: \cite{vanderwalt09}; \cite{vanderwalt11}), 
a stellar pulsation (\cite{inayoshi13}; \cite{sanna15}), 
a circumbinary accretion disk \citep{araya10}, 
and a rotation of spiral shocks within a gap region in a circumbinary disk \citep{parfenov14}. 
The first model is based on changes in the flux of seed photons, 
while the remaining three ones are based on changes in the temperature of dust grains 
at the masing regions. 

Long-term and frequent flux monitoring 
was conducted toward source samples only less than 20{\%}, 
which was estimated from $\sim$200 sources 
(e.g., \cite{goedhart04}; Szymczak et al.\ \yearcite{szymczak15}; \cite{maswan16}) 
in more than 1000 methanol masers samples 
(e.g., \cite{pestalozzi05}; \cite{xu09}; 
\cite{caswell10}, \yearcite{caswell11}; \cite{green10}, \yearcite{green12b}; 
\cite{olmi14}; \cite{sun14}; \cite{breen15}, and references therein). 
Furthermore, most of the monitored sources have been selected 
simply by their peak flux densities $>$ 5~Jy in \citet{szymczak15}. 
We initiated a long-term, highly frequent, and unbiased monitoring 
project using the Hitachi 32-m radio telescope \citep{yonekura16} 
on 30 December 2012 
toward a large sample of the 6.7~GHz methanol masers (442 sources) 
that are located in declination $>$ \timeform{-30D}. 
The observations have been carried out daily, monitoring a spectrum 
of each source with intervals of 9--10 days. 
The detailed results of this project will be reported in forthcoming papers. 
In this paper, we focus on remarkable flux variability 
detected in the 6.7~GHz methanol masers associated with G~014.23$-$00.50 
(hereafter G~014.23). 

The 6.7~GHz methanol maser emission in G~014.23 
was discovered by the methanol multibeam survey toward the 
galactic longitude \timeform{6-20D} \citep{green10}. 
G~014.23 is located in the infrared dark cloud (IRDC) 
G~14.225$-$00.506, which is first identified by \citet{peretto09} 
as a proto-OB association \citep{povich10}. 
This IRDC, also recognized as M17 SWex \citep{povich10}, 
is part of extended ($\sim80$~pc~$\times$~$20$~pc) 
and massive ($> 10^{5}$~\MO) molecular clouds discovered by 
\citet{elmegreen76} 
and is located at the southwest of the Galactic H\emissiontype{II} region M17. 
The distance to M17 has been estimated 
to be 1.98$^{+0.14}_{-0.12}$~kpc through trigonometric parallaxes 
of the methanol masers \citep{xu11}. 
Systemic velocities of M17 and M17 SWex 
are the same within $\sim$1~km~s$^{-1}$ 
in the line emissions of HCO$^{+}$, N$_2$H$^{+}$ 
($V_{\mathrm{lsr}} \sim 18.5$--19.5~km~s$^{-1}$: \cite{schlingman11}; \cite{shirley13}) 
and the absorption of H$_2$CO 
($V_{\mathrm{lsr}} \sim 19$--20~km~s$^{-1}$: \cite{sewilo04}; \cite{okoh14}). 
Based on these results, we adopt the distance of G~014.23 
to be 2~kpc in this paper. 
The methanol maser emission is detected at ``Hub-N", a denser 
dust region where two filaments intersect within the size of $\sim$1~pc 
\citep{busquet13} and the total mass of $\sim$10$^{3}$~\MO \citep{busquet16}. 
This region exhibits signs of active star formation, 
such as higher temperature in NH$_3$ line emissions ($T_{\mathrm{rot}} \sim 15$~K) 
and larger non-thermal velocity dispersion ($\sigma_{\mathrm{NT}} \sim 1$~km~s$^{-1}$) 
than those of the filaments in other regions 
($T_{\mathrm{rot}} \sim 11$~K, $\sigma_{\mathrm{NT}} \sim 0.6$~km~s$^{-1}$) 
\citep{busquet13}. 

\begin{figure}
 \begin{center}
  \includegraphics[width=8cm]{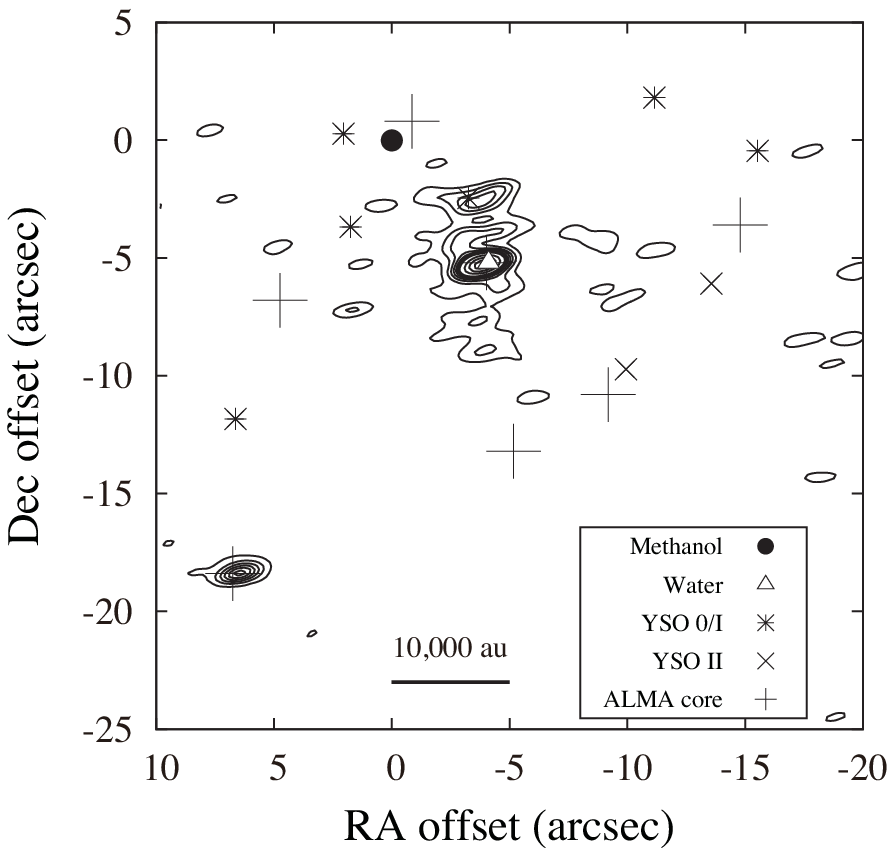} 
 \end{center}
\caption{Close-up map of the 6.7~GHz methanol maser position 
(filled circle) in the dense dust region ``Hub-N" in G~014.23. 
The origin of the map corresponds to the absolute position of the 
methanol maser emission at the local standard of rest (LSR) velocity 
$V_{\mathrm{lsr}} = 25.3$~km~s$^{-1}$ 
[$\alpha$(J2000.0) = \timeform{18h18m12s.59}, 
$\delta$(J2000.0) = \timeform{-16D49'22".8}] 
(\cite{green10}: positional uncertainty of $\sim$\timeform{0".4}). 
The linear spatial scale is shown by the ruler at the bottom. 
The 22.2~GHz water masers (\cite{wang06}: blank triangle) 
and YSOs identified as the evolutionary stage 0/I (asterisk) and II (cross) 
by Povich and Whitney (\yearcite{povich10}) are shown in the map. 
Their positional uncertainties were $\sim$\timeform{0".01} and $\sim$\timeform{0".3}. 
The masers and YSOs are superposed on the contour map of the dust emission 
at 1.3~mm obtained using the SMA \citep{busquet16}, 
which is represented with the same contour levels 
used in the right panel of their figure~2 
(from 3 to 21$\sigma$ in steps of 3$\sigma$, and from 21 to 51$\sigma$ 
in steps of 10$\sigma$, where $\sigma$ is the rms of the map 1~mJy~beam$^{-1}$). 
Their synthesized beam is \timeform{1".46}$\times$\timeform{0".46} 
(position angle of \timeform{-78D}), 
and their positional uncertainties are less than \timeform{1"}. 
The peak positions of pre/proto-stellar cores identified using the ALMA 
at 3~mm \citep{ohashi16} are shown by large crosses 
(the synthesized beam of \timeform{3".7}$\times$\timeform{2".0} ). 
}\label{fig1}
\end{figure}

Among the five-times observations toward G~014.23 by \citet{green10}, 
the methanol masers were detected only twice, being interpreted as 
flaring in flux variations. 
The water maser emission at 22.2~GHz was also detected in the same 
dusty area Hub-N in the IRDC G~14.225$-$00.506 \citep{wang06}. 
Figure~\ref{fig1} shows the methanol and water masers positions 
superposed on the dust emission at $\lambda = 1.3$~mm 
obtained using the submillimeter array (SMA: \cite{busquet16}), 
with the positions of pre/proto-stellar cores identified using the 
atacama large millimeter/submillimeter array (ALMA) 
at $\lambda = 3$~mm \citep{ohashi16} 
and young stellar objects (YSOs) identified by Povich and Whitney (\yearcite{povich10}). 
In the high-resolution observations using the SMA, 
the water masers are presented as being associated 
with the brightest dust condensation MM1a 
with the mass of $\sim$13~\MO, 
while in the ALMA observation 
the methanol masers are associated with a proto-stellar core 
with the mass of $\sim$13~\MO. 
The proto-stellar core includes an intermediate-mass YSO 
G014.2300$-$00.5097 (\cite{povich10}: $\sim 10^{2}$~\LO, 3.3~\MO). 
This intermediate-mass YSO is classified as the stage 0/I 
according to the definition by \citet{robitaille06}, 
suggesting that infalling activities from a surrounding envelope are still dominant. 
Furthermore, an excess in 4.5~$\micron$ band is also detected 
at this intermediate-mass YSO, and it might be 
due to unresolved analogs of the extended green objects, 
which are candidates of active outflows ejected from high-mass YSOs at 
an early evolutionary phase \citep{cyganowski08}. 
These characteristics imply that this intermediate-mass YSO accompanied by 
the methanol masers is a precursor of a B-type star 
on the main sequence \citep{povich10}. 

In this paper, we discuss remarkable flux variability 
detected in the 6.7~GHz methanol masers of G~014.23. 
The structure of this paper is as follows: 
in sections \ref{section2} and \ref{section3}, 
we present observations and results with data reduction 
for determination of periods, respectively. 
In section \ref{section4}, we discuss suitable models 
to explain the remarkable flux variability.

\section{Observations}\label{section2}
\subsection{Periodic monitoring at 10-day intervals}\label{section2.1}

\begin{table}
  \tbl{Time range of the periodic flux monitoring 
  using the Hitachi 32-m radio telescope.}{%
  \begin{tabular}{lcccc}
      \hline
      Period & \multicolumn{2}{c}{Time range of monitoring} \\ \cline{2-3} 
                   & A.D. & MJD \\
                   & [yyyy/mm/dd] & [day] \\
      \hline
\multicolumn{3}{l}{\textbf{Periodic monitoring at 10-day intervals}} \\
 & 2013/01/05 -- 2014/01/10 & 56297 -- 56667 \\ 
\multicolumn{3}{l}{\textbf{Daily monitoring}} \\
 & 2014/05/07 -- 2016/01/21 & 56784 -- 57408 \\ 
\multicolumn{3}{l}{\textbf{Intraday monitoring}} \\
0 & 2014/03/29 -- 2014/05/01 & 56745 -- 56778 \\ 
I & 2015/08/09 -- 2015/08/19 & 57243 -- 57253 \\ 
II & 2015/09/01 -- 2015/09/16 & 57266 -- 57281 \\ 
III & 2015/09/30 -- 2015/10/04 & 57295 -- 57299 \\ 
IV & 2015/10/26 -- 2015/10/31 & 57321 -- 57326 \\ 
\hline
    \end{tabular}}\label{tab1}
\begin{tabnote}
Column~1: Identification number for intraday monitoring; 
Columns~2, 3: Time range of each periodic flux monitoring 
as A.D. and modified julian day. 
\end{tabnote}
\end{table}

We made flux monitoring with intervals of 9--10 days, 
the time range of which is listed in table~\ref{tab1}, 
using the Hitachi 32-m radio telescope 
toward the target source G~014.23. 
The full-width at half maximum (FWHM) of the beam is \timeform{4.6'} 
and the pointing accuracy is better than \timeform{30"}. 
Each observation was made at almost the same 
azimuth and elevation angles 
(\timeform{169D}, \timeform{35D}) to minimize the effects of 
pointing errors for observed flux densities. 
Left-circular polarization (LCP) signals were recorded on hard disk drive 
by using the 64~Mbps mode (16 mega-samples per second 
with 4 bit sampling) through K5/VSSP32 sampler 
and converted to spectra using software spectroscopic system 
(Nitsuki 9270: \cite{kondo08}). 
The bandwidth is 8~MHz (RF: 6664--6672~MHz), 
covering the velocity range of $\sim$360~km~s$^{-1}$. 
The number of frequency channels is originally 2,097,152, 
but they are bunched into 8,192 channels, 
yielding the velocity resolution of 0.044~km~s$^{-1}$. 
The system noise temperature 
at the zenith after the correction for the atmosphere opacity 
($T_{\mathrm{sys}}^{\ast}$) 
and the aperture efficiency ($\eta_{\mathrm{A}}$) 
is typically 30--40~K and 55--75{\%} at 6.7~GHz, respectively. 
The rms noise level (1$\sigma$) is achieved to be $\sim$0.3~Jy 
with an integration time of 5~min 
and further 3~ch smoothing, 
yielding the velocity resolution of 0.13~km~s$^{-1}$. 
The antenna temperature 
($T_{\mathrm{a}}^{\ast}$) was measured by the chopper-wheel method. 
We observed a sky area 
(\timeform{+60'} offset in right ascension direction) 
as off-source data just after each on-source scan 
in the $T_{\mathrm{a}}^{\ast}$ measurement.

\subsection{Daily monitoring}\label{section2.2}

We initiated daily monitoring on 7 May 2014 (modified julian day (MJD) $=$ 56784) 
with the same observational setup as that described in section~\ref{section2.1}. 
The time range of the daily monitoring is listed in table~\ref{tab1}. 
During this observation period, 
there are blank dates when the daily monitoring was not made 
due to system maintenance 
or other observations as follows: 
9--20 June and 12--27 July 2014 
(MJD 56817--56828 and 56850--56865, respectively), 
and 25 May to 6 June and 20--31 August 2015 
(MJD 57167--57179 and 57254--57265, respectively). 
The stability of the system was evaluated by 
daily monitoring of the 6.7~GHz methanol masers in G~012.02$-$00.03 
and G~069.52$-$00.97 (Onsala~1) in the same periods. 
Their peak flux densities were stronger than 100~Jy 
and stable whose variations were smaller than 10{\%} during the periods. 
The variations were estimated from 
the modulation index expressed as the standard deviation 
normalized by the averaged value of flux densities in each source. 
As a result, the stability of the observational system is estimated 
to be better than 15{\%}.

\subsection{Intraday monitoring}\label{section2.3}

We carried out intraday monitoring in five epochs (table~\ref{tab1}) 
with the same observational setup as that described in section~\ref{section2.1}, 
except for the azimuth and elevation angles. 
The intraday monitoring is classified into the periods 0 and I--IV; 
the former was made before an evaluation of the periodicity 
while the latter were made after the evaluation. 
In the period 0 in 2014, 
we focused on investigating whether there occurred flux variations 
weaker than 0.9~Jy, which was equal to the detection limit (3$\sigma$) 
with the integration time of 5~min. 
The observational cycle of 10.5~min that consists of 5~min 
for the target source, half a minute for slewing, and 5~min for the off source 
was repeated during the elevation angles $\gtrsim$~\timeform{30D} until 27 April 
(MJD 56774) and $\geq$~\timeform{15D} from 28 April (MJD 56775). 
The detection limits of 3$\sigma =$~0.15--0.20~Jy were achieved 
in each observational day by integrating all the scans. 
On the other hand, in the periods I--IV in 2015, 
we focused on measuring accurate time-scale of rises and decays 
of flux densities and profiles of flux variations during each period. 
The maser G~014.10$+$00.08 was used as a calibrator, 
located close to the target with the separation angle of \timeform{36.6'}. 
Its peak flux densities were 70--80~Jy 
and stable whose variation was smaller than 10{\%} during each period. 
The cycle consisting of $\sim$11~min single-point observation 
of G~014.23 (5~min on-source and 5~min off-source) 
and $\sim$15~min nine-point cross-scan of G~014.10$+$00.08 
was repeated during the elevation angles $\geq$~\timeform{15D} 
to calibrate pointing errors and atmospheric effects. 
Solutions for the calibration obtained within two to three days at the beginning 
of each period were applied to all the data during the periods.

\section{Results}\label{section3}

\begin{figure}
 \begin{center}
  \includegraphics[width=8cm]{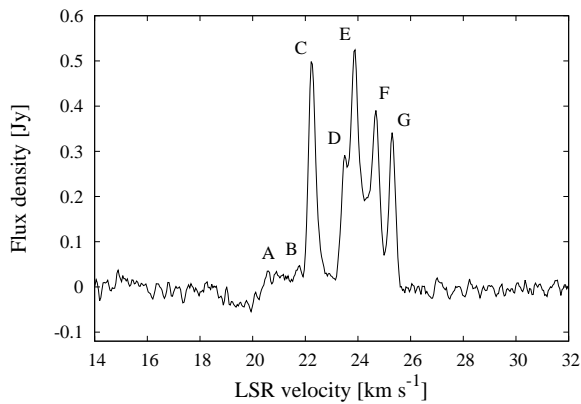} 
 \end{center}
\caption{
Average spectrum of the 6.7~GHz methanol maser emission in G~014.23. 
All 504 scans of 5~min duration obtained 
at the same azimuth (\timeform{169D}) and elevation (\timeform{35D}) 
angles were used. 
This average was weighted by $\sigma^{-2}$ (inverse square of rms noise level), 
and the 3$\sigma$ detection limit of 0.039~Jy was achieved. 
Labels A, B, C, D, E, F, and G indicate spectral components 
at $V_{\mathrm{lsr}} =$ 20.98, 21.75, 22.30, 23.46, 23.91, 24.69, 
and 25.30~km~s$^{-1}$, respectively. 
Note that the components A and B do not fulfill the 3$\sigma$ criterion to be real 
in this figure. 
}\label{fig2}
\end{figure}

First of all, we made 
an average spectrum of G~014.23 by integrating 
all observational data to identify spectral components, as shown in figure~\ref{fig2}. 
This average was weighted by $\sigma^{-2}$, an inverse square of rms noise level. 
From the intraday monitoring described in section~\ref{section2.3}, 
we picked up one 5-min scan for each day 
observed at the same azimuth and elevation angles (\timeform{169D}, \timeform{35D}) 
as the periodic monitoring in section~\ref{section2.1} and \ref{section2.2}. 
This procedure provides us with a chance for handling all the data 
equally without being biased toward data during flares. 
As a result, the average spectrum was obtained from 
504 scans of 5~min duration, or 
42~hrs (2,520~min) integration time in total, 
achieving the detection limit as 3$\sigma$ of 0.039~Jy. 
In this spectrum, 
five spectral components were identified at 
$V_{\mathrm{lsr}} =$ 22.30, 23.46, 23.91, 24.69, 
and 25.30~km~s$^{-1}$. 
In addition, 
two spectral components at 
$V_{\mathrm{lsr}} =$ 20.98 and 21.75~km~s$^{-1}$ 
were identified in limited periods of the monitoring (see section \ref{section3.1}). 
These are labeled as A, B, C, D, E, F, and G, respectively 
in order of velocity. 
The six components A--F are newly detected. 
The significant dip of $\sim$0.05~Jy was detected 
at $V_{\mathrm{lsr}} \sim$19--20~km~s$^{-1}$ 
and could be an absorption line because of 
$V_{\mathrm{lsr}}$ close to one of H$_2$CO absorption line \citep{sewilo04}. 
However, it is beyond the scope of this paper.

\subsection{Flux variability}\label{section3.1}

\begin{figure*}
 \begin{center}
  \includegraphics[width=16cm]{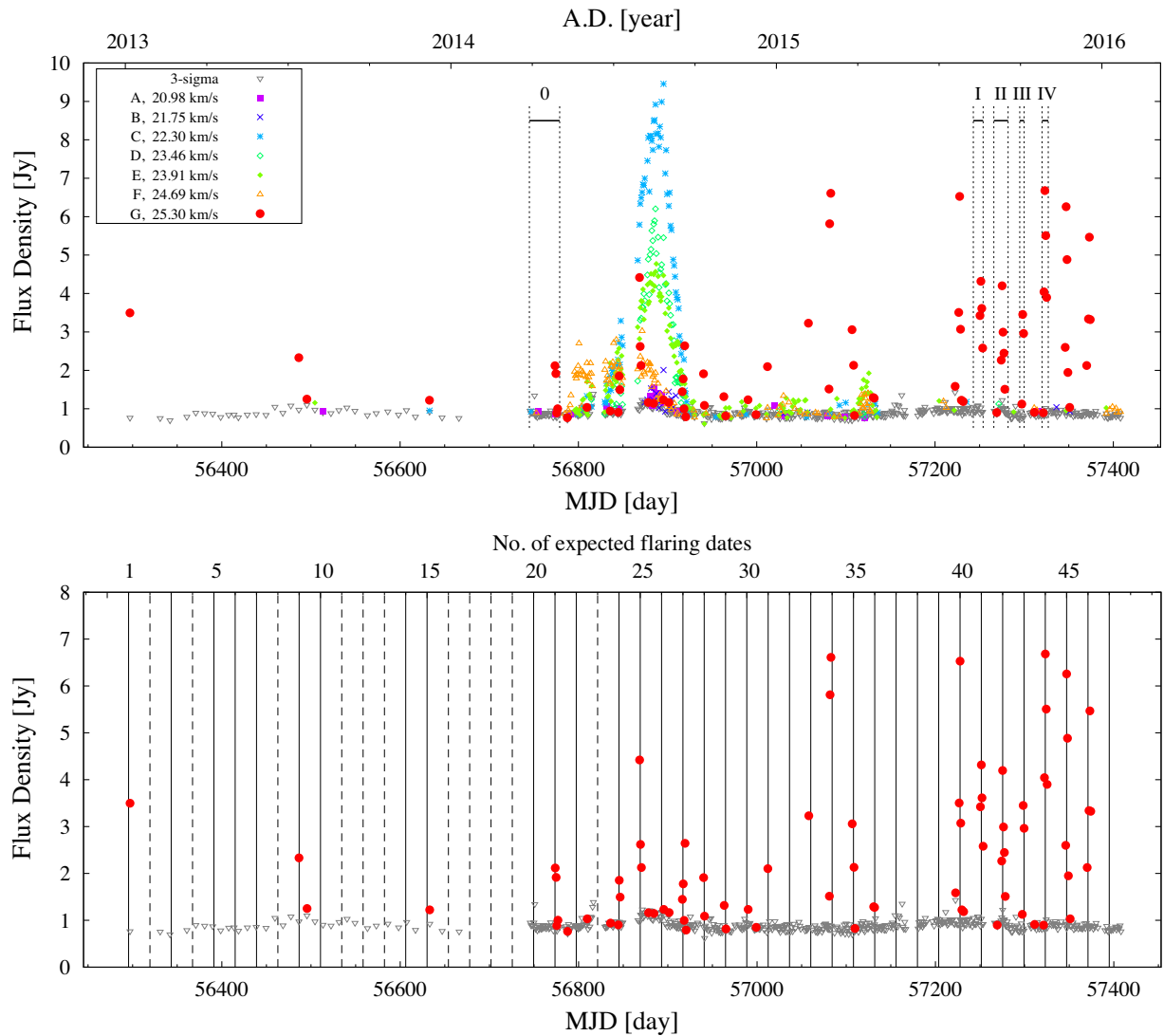} 
 \end{center}
\caption{
Upper-panel: Flux variation of each spectral component of the 6.7~GHz methanol 
masers in G~014.23 from 5 January 2013 to 21 January 2016 
(MJD 56297--57408). 
Each symbol shows the spectral components referred as legends at top-left corner. 
The time intervals denoted by dotted vertical lines correspond to the periods 
of the intraday flux monitoring 0--IV, 
as described in section~\ref{section2.3}. 
Lower-panel: Same as the upper-panel but only for the component G. 
Solid and dashed lines show the expected flaring dates 
with and without observational data, 
which are calculated in section~\ref{section3.2} and listed in table~\ref{tab2}. 
}\label{fig3}
\end{figure*}

Figure~\ref{fig3} shows time variations of flux densities of all spectral 
components from 5 January 2013 to 21 January 2016 (MJD 56297--57408). 
Concerning the data points during the intraday monitoring, 
only the results measured at the same azimuth and elevation angles 
as the periodic monitoring are shown in figure~\ref{fig3}. 

\begin{figure}
 \begin{center}
  \includegraphics[width=8cm]{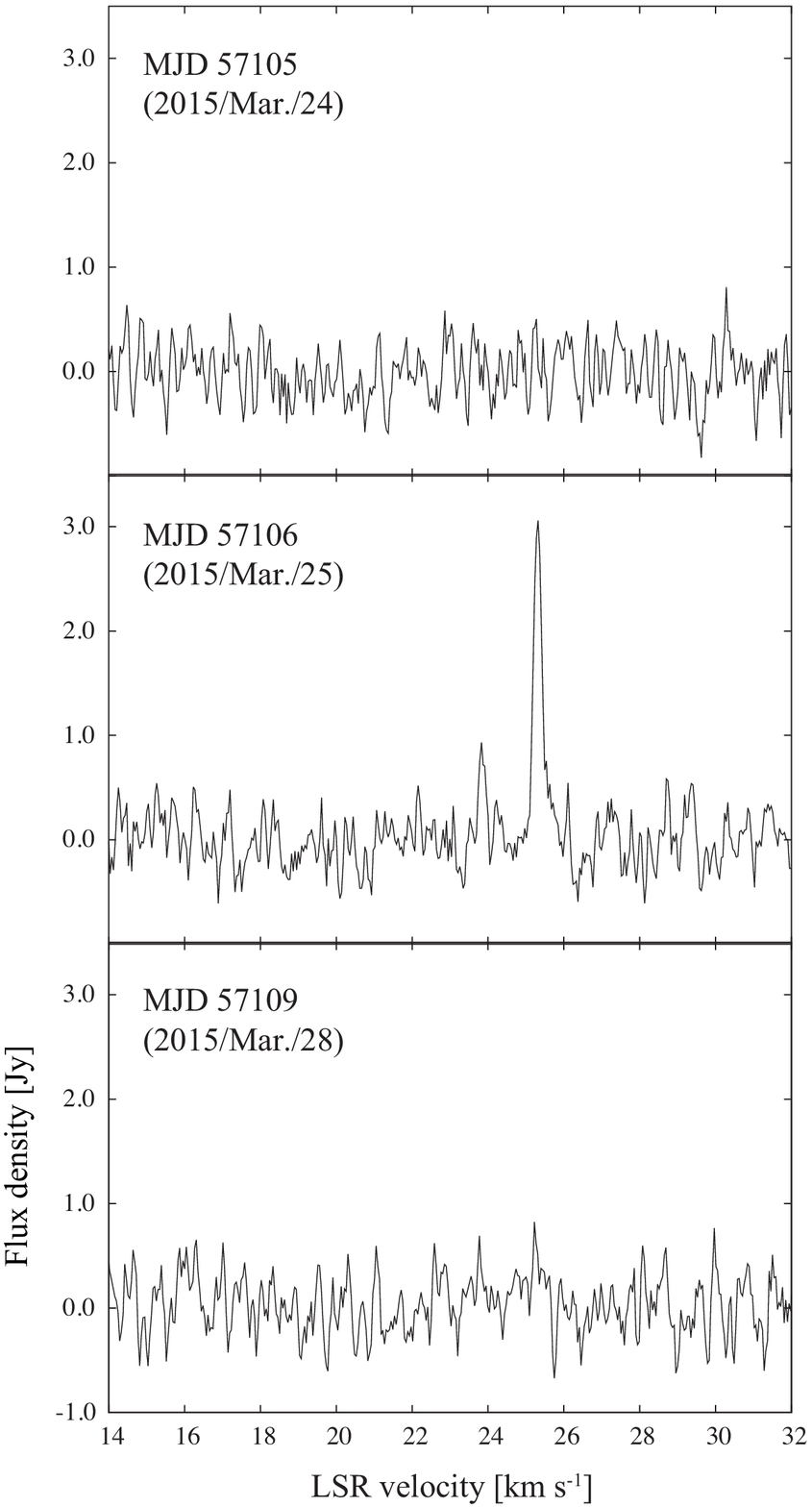} 
 \end{center}
\caption{
Typical flaring activity detected toward the spectral component G. 
Top, middle, and bottom panel shows the maser spectrum 
observed on MJD 57105, 57106, and 57109 (24, 25, and 28 March 2015), 
respectively. 
}\label{fig4}
\end{figure}

The component G showed remarkable variability (filled circle in figure~\ref{fig3}). 
The flux density of this component in a quiescent phase 
is usually below the detection limit of $\sim$0.9~Jy. 
This component, however, sometimes rises above the detection limit. 
The time-scale of the rise is two days or shorter in most cases, 
and the component decays within a few days, which could be called a ``flare." 
Figure~\ref{fig4} shows representative spectra 
as a typical case for the flaring activities. 
There was no emission in the top panel observed on MJD 57105, 
and the component G was detected on the next day (MJD 57106) 
shown in the middle panel, 
while it disappeared three days later (MJD 57109) in the bottom panel. 
We have detected clear periodicity in this component, 
as described in detail in section~\ref{section3.2}. 

The components A--F emerged on MJD 56788 (11 May 2014) 
and showed flux variations until MJD 57130 (18 April 2015). 
In particular, three components C, D, and E 
(asterisk, blank diamond, and filled diamond in the upper-panel 
of figure~\ref{fig3}) 
showed notable flux variations until MJD 56943, 
which were brighter than the maximum flux density 
of the flaring component G. 
The time-scale of the variations was longer than 100~days. 
No correlation was seen 
between the long-term flux variability of the components A--F 
and the flaring activities of the component G. 
The maximum flux density of the component G 
during our monitoring was detected 
after the period when the components A--F were bright, 
suggesting also that those have poor correlation. 
In this paper, 
we mainly focus on the flaring activities of the component G.

\subsection{Periodicity evaluated by Lomb-Scargle periodogram}\label{section3.2}

\begin{figure}
 \begin{center}
  \includegraphics[width=8cm]{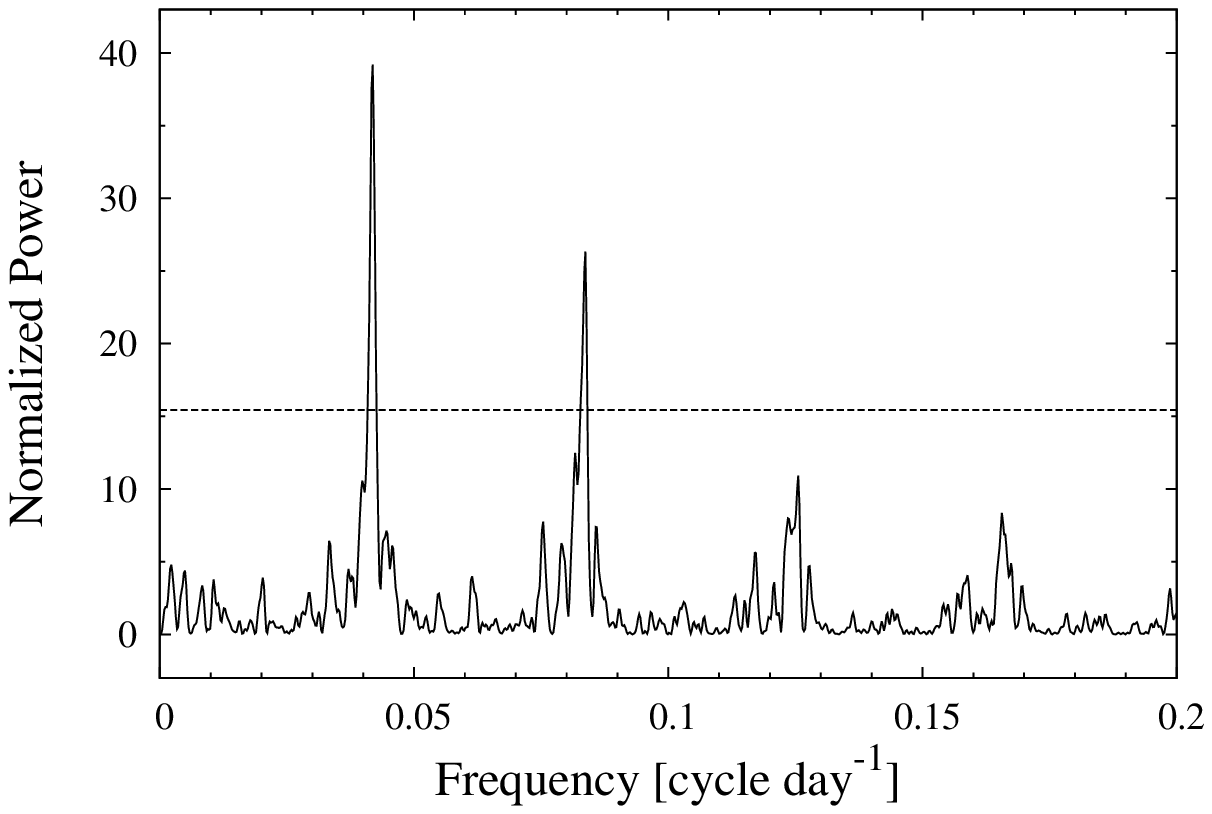} 
 \end{center}
\caption{
L-S periodogram applied to the flaring spectral component G. 
A false-alarm probability $\leq 10^{-4}$, corresponding to 
the power level of 15.4, is denoted by a dashed horizontal line. 
The peak detected at the frequency around 0.084~cycle~day$^{-1}$ 
with the power level of 26.3 must be a harmonic signal of the derived 
period as 23.9~days. 
}\label{fig5}
\end{figure}

We adopted the Lomb-Scargle (L-S) periodogram (\cite{lomb76}; \cite{scargle82}) 
to evaluate periodicity for the flaring activities of the component G. 
This is the most reliable method 
to search for periodicity in flux variations of the methanol masers 
\citep{goedhart14}. 
In this paper, we regard an oversampling factor of 4
and frequencies with false-alarm probability $\leq 10^{-4}$ as significant, 
which are the same manner as \citet{goedhart14}. 
The L-S periodogram adopted to all the data in MJD 56297--57408 
is shown in figure~\ref{fig5}, 
with the power level of 15.4 in the dashed line that corresponds to 
the false-alarm probability of $10^{-4}$. 
To avoid any bias, we used one 5-min scan for each day during 
the intraday monitoring, 
measured at the same azimuth and elevation angles 
as the other monitoring in section~\ref{section2.1} and \ref{section2.2}. 
We evaluated the period of 23.9~days as the significant detection 
(the power level of 39.2), corresponding to the frequency of 
$\sim$0.0419~cycle~day$^{-1}$. 
We evaluated calculation errors to be $\pm 0.1$~days 
from the frequency resolution in the L-S periodogram. 

The derived period of 23.9~days 
was also verified from comparison 
between observed flaring dates MJD$_{\mathrm{obs}}$ 
and expected ones from the periodicity. 
The expected flaring dates MJD$_{\mathrm{calc}}$ were calculated by 
$\mathrm{MJD}_{\mathrm{calc}} = \mathrm{MJD}_{0} + 23.9 \cdot n$ [day], 
where $n$ is an integer, and MJD$_{0} = 57275.4$ 
is the reference date estimated from the period II of the intraday monitoring. 
In the period II, it is difficult to define 
the accurate time when the component G reached its maximum on MJD 57275 
(see figure~\ref{fig8}b: relative MJD $=$ 2), 
causing errors of $\pm 0.5$~days 
at most in the flaring peak timing. 
The expected flaring dates for 47 flares 
are shown as solid and dashed lines in the lower panel of figure~\ref{fig3} 
and summarized in table~\ref{tab2}. 
Until MJD 56745, 
spectra were obtained every 9--10~days, and 
thus no observations were executed within three days of MJD$_{\mathrm{calc}}$ 
for No. 2, 4, 8, 11--13, and 16. 
Also for the expected flares of No. 17--19 and 23, 
no observations were executed within three days of MJD$_{\mathrm{calc}}$ 
due to unavailability of the telescope. 
These are denoted by the symbol ``$\cdots$" in column~3 of table~\ref{tab2}. 
By excluding these expected flares, we compared 36 flares to evaluate 
whether the derived period is reasonable. 
As a result, 67{\%} (24/36) of the observed flares 
coincide with the expected flaring dates within three~days 
($| \mathrm{MJD}_{\mathrm{obs}} - \mathrm{MJD}_{\mathrm{calc}} | \leq 1$~day: 14 flares, 
$1 < | \mathrm{MJD}_{\mathrm{obs}} - \mathrm{MJD}_{\mathrm{calc}} | \leq 2$~days: 6 flares, 
and $2 < | \mathrm{MJD}_{\mathrm{obs}} - \mathrm{MJD}_{\mathrm{calc}} | \leq 3$~days: 4 flares). 
On the other hand, no flares above 3.5$\sigma$ were detected at the dates 
out of three~days of MJD$_{\mathrm{calc}}$. 
The fact that the observed flares do not necessarily coincide with the expected flaring dates 
would not be attributed to the errors of 0.1~days in the L-S periodogram calculation 
and 0.5~days in the definition of the reference date MJD$_{0}$. 
No correlation was seen 
between $| \mathrm{MJD}_{\mathrm{obs}} - \mathrm{MJD}_{\mathrm{calc}} |$ 
and the periodic cycle $n$. 
The not necessarily coincidence can be attributed to the following two reasons: 
(1)~Intrinsic difference in the time-scale of flux rising in the flaring activities 
may cause the difference of the observed flaring dates from the expected ones. 
Some flares might show the rising time over two days, one of which was observed 
in the period I of the intraday monitoring (see section~\ref{section3.3}). 
The peak timing of flares, therefore, might differ 
even in the case that the start timing of flares is same. 
(2)~No observations in one day earlier or later than MJD$_{\mathrm{calc}}$, 
occurred even in the observational dates during the daily monitoring 
from MJD~56784 (No. 25, 26, 28--35, 37, 38, and 47), 
may cause 1~day error in the difference between observed and expected 
flaring dates. 

\begin{figure}
 \begin{center}
  \includegraphics[width=8cm]{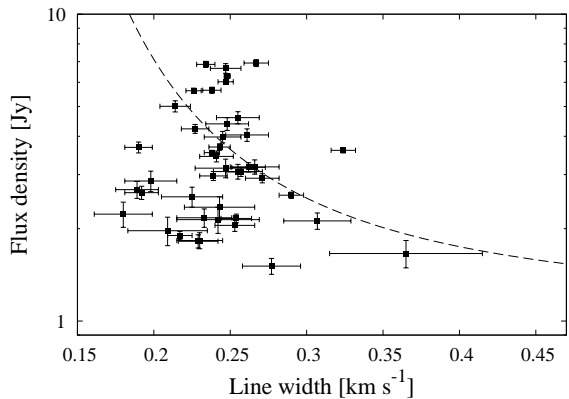} 
 \end{center}
\caption{
Relationship between the flux density and line width of the flaring 
component G. 
Filled squares represent the observed values showing flux densities 
brighter than the 5$\sigma$ detection of $\sim$1.5~Jy. 
A dashed line shows the best-fit result of the fitting; 
$\log F = A + B \cdot \Delta v^{-2}$, where the parameter $B$ is 
$0.033 \pm 0.002$ while the parameter $A$ is fixed as the upper limit 
in a quiescent spectrum of 0.039~Jy (see section \ref{section3.4}). 
}\label{fig6}
\end{figure}

Non-detection might be attributed to an intrinsic property of the flaring 
component G. 
In the cases of 
$| \mathrm{MJD}_{\mathrm{obs}} - \mathrm{MJD}_{\mathrm{calc}} | \leq 1$~day, 
such as No. 7 and No. 14 in table~\ref{tab2}, 
the non-detection is likely due to the absence of flare. 
The absence of flare is more strongly suggested in the period of more frequent 
observations (No. 20--47). 
The variability of flare amplitude 
might be due to a saturation level of the flaring maser gas. 
The time variation of the line width of the flaring component G implies 
an anti-correlation with the flux density as shown in figure~\ref{fig6}. 
The line width in figure~\ref{fig6} is estimated as FWHM by the gaussian fitting, 
and the flare components brighter than the 5$\sigma$ detection of $\sim$1.5~Jy 
are plotted, which are well fitted within errors of 10{\%}. 
According to model calculations, the line width tends to be narrower as its intensity 
increases, as long as the maser is unsaturated \citep{goldreich74}. 
Under the condition of unsaturation, the logarithm of the flux density, 
$\log F$, is predicted to be proportional to the inverse square of the line width, 
$\Delta v^{-2}$. 
The best-fit result using $\log F = A + B \cdot \Delta v^{-2}$ 
is shown in figure~\ref{fig6}, where the parameter $B$ is 
$0.033 \pm 0.002$ while the parameter $A$ is fixed as the upper limit 
in a quiescent spectrum of 0.039~Jy (see section \ref{section3.4}). 
Our results may suggest that the component G in the flaring phase is still unsaturated. 
The unsaturated state might cause the variability of flare amplitude 
and result in the non-detection of some flares, 
whose peak flux density should be lower than the detection limit. 
For instance, No. 20 flare having very weak peak flux density of 0.57~Jy 
observed in the period 0 of the intraday monitoring 
cannot be detected with the 3$\sigma$ detection limit of 0.9~Jy 
with the integration time of 5~min in the periodic monitoring 
(see section~\ref{section3.3}).

We conclude that the flaring activities of the spectral component G 
have regularly occurred with 
the period of 23.9$\pm 0.1$~days at least in 47 cycles, 
corresponding to $\sim$1,100~days. 
The derived period of 23.9~days is the shortest one observed 
in the masers at around high-mass YSOs so far, 
compared to 29.5~days in methanol \citep{goedhart09}, 
$\sim$2~yrs in silicon monoxide \citep{ukita81}, 
237~days in formaldehyde \citep{araya10}, 
25--30~days in hydroxyl \citep{green12a}, 
and 34.4~days in water \citep{szymczak16}.

\renewcommand{\arraystretch}{1.1}
\begin{table}
  \tbl{Comparison between observed and expected flaring dates 
for each flare in 47 periodic cycles.}{
\footnotesize
  \begin{tabular}{lrcccc}
      \hline
      No. & \multicolumn{1}{c}{$n$} & MJD$_{\mathrm{obs}}$ 
             & $F_{\mathrm{p}}$ & MJD$_{\mathrm{calc}}$ & Difference \\ 
             &         & [day]    & [Jy]     & [day]    & [day] \\
      \hline
\hspace{1.4mm}1 & $-$41 & 56297.0 & 3.5 & 56295.5 & $+$1.5  \\
\hspace{1.4mm}2 & $-$40 & $\cdots$ &  & 56319.4 &  \\
\hspace{1.4mm}3 & $-$39 & -- & $< 0.70$ & 56343.3 & -- \\
\hspace{1.4mm}4 & $-$38 & $\cdots$ &  & 56367.2 &  \\
\hspace{1.4mm}5 & $-$37 & -- & $< 0.87$ & 56391.1 & -- \\
\hspace{1.4mm}6 & $-$36 & -- & $< 0.85$ & 56415.0 & -- \\
\hspace{1.4mm}7 & $-$35 & -- & $< 0.86$ & 56438.9 & -- \\
\hspace{1.4mm}8 & $-$34 & $\cdots$ &  & 56462.8 &  \\
\hspace{1.4mm}9 & $-$33 & 56486.5 & 2.3 & 56486.7 & $-$0.2  \\
10 & $-$32 & -- & $< 0.90$ & 56510.6 & -- \\
11 & $-$31 & $\cdots$ &  & 56534.5 &  \\
12 & $-$30 & $\cdots$ &  & 56558.4 &  \\
13 & $-$29 & $\cdots$ &  & 56582.3 &  \\
14 & $-$28 & -- & $< 0.96$ & 56606.2 & -- \\
15 & $-$27 & 56633.1 & 1.2 & 56630.1 & $+$3.0 \\
16 & $-$26 & $\cdots$ &  & 56654.0 &  \\
17 & $-$25 & $\cdots$ &  & 56677.9 &  \\
18 & $-$24 & $\cdots$ &  & 56701.8 &  \\
19 & $-$23 & $\cdots$ &  & 56725.7 &  \\
20 & $-$22 & 56749.8 & \hspace{2.9mm}0.57$^{\ast}$ & 56749.6 & $+$0.2  \\
21 & $-$21 & 56773.8 & 2.1 & 56773.5 & $+$0.3  \\
22 & $-$20 & -- & $< 0.90$ & 56797.4 & -- \\
23 & $-$19 & $\cdots$ &  & 56821.3 &  \\
24 & $-$18 & 56845.6 & 1.9 & 56845.2 & $+$0.4  \\
25 & $-$17 & 56868.5 & 4.4 & 56869.1 & \hspace{1.4mm}$-$0.6$^{\dagger}$  \\
26 & $-$16 & 56895.4 & 1.2 & 56893.0 & \hspace{1.4mm}$+$2.4$^{\dagger}$  \\
27 & $-$15 & 56917.4 & 1.8 & 56916.9 & $+$0.5  \\
28 & $-$14 & 56940.3 & 1.9 & 56940.8 & \hspace{1.4mm}$-$0.5$^{\dagger}$  \\
29 & $-$13 & 56963.2 & 1.3 & 56964.7 & \hspace{1.4mm}$-$1.5$^{\dagger}$  \\
30 & $-$12 & 56990.2 & 1.2 & 56988.6 & \hspace{1.4mm}$+$1.6$^{\dagger}$  \\
31 & $-$11 & 57012.1 & 2.1 & 57012.5 & \hspace{1.4mm}$-$0.4$^{\dagger}$  \\
32 & $-$10 & -- & $< 0.91$ & 57036.4 & \hspace{1.4mm}--$^{\dagger}$ \\
33 & $-$9 & 57058.0 & 3.2 & 57060.3 & \hspace{1.4mm}$-$2.3$^{\dagger}$  \\
34 & $-$8 & 57082.9 & 6.6 & 57084.2 & \hspace{1.4mm}$-$1.3$^{\dagger}$  \\
35 & $-$7 & 57106.8 & 3.1 & 57108.1 & \hspace{1.4mm}$-$1.3$^{\dagger}$  \\
36 & $-$6 & 57130.8 & 1.3 & 57132.0 & $-$1.2  \\
37 & $-$5 & -- & $< 0.90$ & 57155.9 & \hspace{1.4mm}--$^{\dagger}$ \\
38 & $-$4 & -- & $< 0.82$ & 57179.8 & \hspace{1.4mm}--$^{\dagger}$ \\
39 & $-$3 & -- & $< 0.91$ & 57203.7 & -- \\
40 & $-$2 & 57227.5 & 6.5 & 57227.6 & $-$0.1  \\
41 & $-$1 & 57251.4 & 4.3 & 57251.5 & $-$0.1  \\
42$^{\ddagger}$ & 0 & 57275.4 & 4.2 & 57275.4 & \hspace{2.3mm}0.0  \\
43 & $+$1 & 57298.3 & 3.5 & 57299.3 & $-$1.0  \\
44 & $+$2 & 57323.2 & 6.7 & 57323.2 & \hspace{2.3mm}0.0  \\
45 & $+$3 & 57347.2 & 6.3 & 57347.1 & $+$0.1  \\
46 & $+$4 & 57373.1 & 5.5 & 57371.0 & $+$2.1  \\
47 & $+$5 & -- & $< 0.76$ & 57394.9 & \hspace{1.4mm}--$^{\dagger}$ \\
\hline
    \end{tabular}}\label{tab2}
\begin{tabnote}
Column~1: Identification number of each flare; 
Column~2: Periodic cycle of each flare from the reference flare No. 42 
(the integer used in section~\ref{section3.2}); 
Columns~3, 4: Observed flaring date and flux density. 
``$\cdots$" denotes that no observations were executed 
within three days of the expected flaring date, 
while ``--" shows no detection of the flares.; 
Column~5: Expected flaring date; 
Column~6: Difference between observed and expected flaring dates.\\ 
Note. -- $^{\ast}$Detected thanks to high-sensitivity 
as the detection limit 3$\sigma$ of 0.19~Jy 
achieved by the integration of all the data in MJD 56750 
(see section~\ref{section3.3}); 
$^{\dagger}$there are no observations in one day earlier or later 
than MJD$_{\mathrm{calc}}$ during the daily monitoring from MJD~56784 
(see section~\ref{section3.2}); 
$^{\ddagger}$reference date for the calculation of the expected flaring dates. 
\end{tabnote}
\end{table}
\renewcommand{\arraystretch}{1}

\subsection{Flux variability in the intraday monitoring}\label{section3.3}

\begin{figure}
 \begin{center}
  \includegraphics[width=8cm]{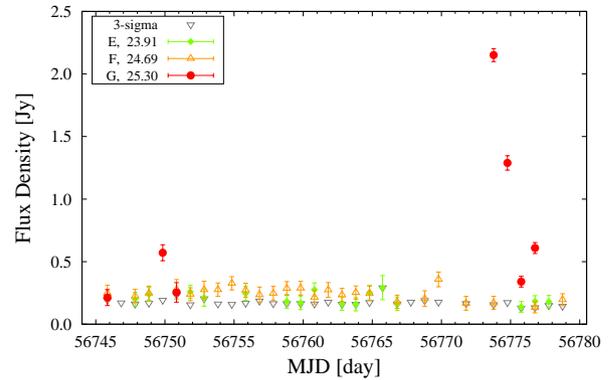} 
 \end{center}
\caption{
Flux variation detected in the period 0 of the intraday monitoring. 
Each point shows the data obtained by the integration of all the data 
in each observational date. 
Each symbol showing each spectral component is unified with one 
used in figure~\ref{fig3}. 
The error bar represents the standard deviation in each date. 
}\label{fig7}
\end{figure}

We present flux variability obtained through the intraday monitoring 
described in section~\ref{section2.3}. 
In the period 0, 
the flaring activities were detected twice in the component G, 
as shown in figure~\ref{fig7}. 
Each point shows the data obtained by the integration of all the data 
in each observational date. 
The integration time is 
$\sim$1.8~hrs (110~min) and $\sim$3.3~hrs (195~min) 
in MJD 56745--56774 and 
56775--56778, respectively, 
yielding typical detection limits (3$\sigma$) of 0.15--0.20~Jy. 
The first flare occurred on MJD 56750 and 56751. 
On MJD 56750, the flux density was reached to the maximum of 0.57~Jy 
with the signal to noise ratio of $\sim$9. 
It was not detected in the periodic monitoring 
with the integration time of 5~min that yielded 
the detection limit (3$\sigma$) of $\sim$0.9~Jy. 
This flaring component decayed to 0.25~Jy, which was 
close to the detection limit, on MJD 56751, 
and disappeared on the next day (MJD 56752). 
The second flare occurred on MJD 56774--56777. 
The peak flux density was 2.2~Jy, and an e-folding time for decay 
was estimated to be 1.3~days. 

\begin{figure*}
 \begin{center}
  \includegraphics[width=16cm]{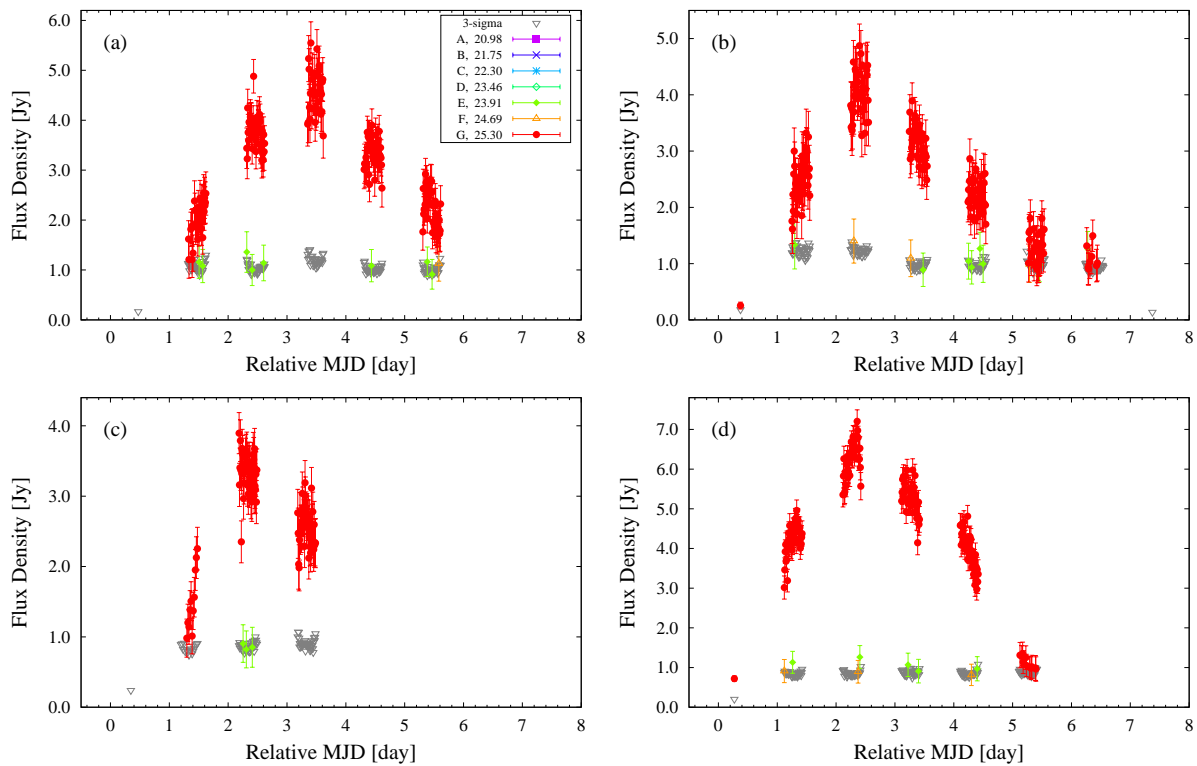} 
 \end{center}
\caption{
Same as figure~\ref{fig7} but for the periods I--IV 
shown in panels (a)--(d), respectively. 
The horizontal axis is MJD relative to the reference date of 
57248, 57273, 57296, and 57321 in each panel, respectively. 
The reference date is defined as the previous day from the date 
when the flaring component is detected in one 5-min scan 
for the first time in each period. 
In the case of no detection in all 5-min scans through 
each observational date, 
an integration with all the data in each date 
was performed to obtain better sensitivities, yielding the detection limit 
(3$\sigma$) of $\sim$0.15--0.20~Jy. 
}\label{fig8}
\end{figure*}

Figure~\ref{fig8}(a)--(d) show results obtained in the periods I--IV 
of the intraday monitoring, respectively. 
The flare of the component G was detected in all the periods. 
In the period I, the rising time to the peak flux density of $\sim$5~Jy 
was two days, and almost same days were 
spent for decaying, showing a symmetric profile of flux variability. 
On the other hand, 
asymmetric profiles were obtained in the periods II and IV. 
In the periods II and IV, 
the peak flux densities were $\sim$4.5 and 7.0~Jy. 
The rising time was two days, 
while the decay has the e-folding time of 2.7 and 2.6~days, respectively. 
In the period III, no decay part of flux variability was observed. 
The rising time to the peak flux density of $\sim$3.5~Jy was within one day, 
while the decaying time seems to be longer judging from figure~\ref{fig8}(c), 
being suggestive of an asymmetric profile. 

Interestingly, we detected the rising time accurately in the period III. 
No emission was detected from an integrated spectrum on MJD 57296 
(relative MJD $=$ 0 in figure~\ref{fig8}c) 
and from the first scan at 4:45:55 UT to the scan at 6:29:35 UT on MJD 57297. 
At the next scan at 7:09:00 UT on MJD 57297, 
the flaring component G was detected for the first time in this period 
above 3$\sigma$ of 0.9~Jy with 5~min integration, 
and then it rose up. 
Although a turning point from rising to decaying was not detected 
on the next day (MJD 57298) since the maximum flux density 
in this period occurred on the first scan 4:27:00 UT, 
the rising time was stringently evaluated to be $\leq$~21.3~hrs (1,278~min). 
This is regarded as an upper limit of the rising time of the flare.

\subsection{Typical spectrum in a quiescent phase}\label{section3.4}

In this section, 
we evaluate a ratio of the peak flux density in the flaring 
and quiescent (non-flaring) phases. 
The quiescent phase MJD$_{\mathrm{q}}$ is defined by 
$\mathrm{MJD}_{\mathrm{q}} = \bigl( \mathrm{MJD}_{0} + 23.9 \cdot \left( n - 0.5 \right) \bigr) \pm 5$ [days], 
where $n$ is an integer, and MJD$_{0}$ 57275.4 is the reference date (see section~\ref{section3.2}). 
We also picked up data taken 
from the intraday periods 0--IV, even in the case of 
being observed at different azimuth and elevation angles 
to obtain the best detection limit. 

\begin{figure}
 \begin{center}
  \includegraphics[width=8cm]{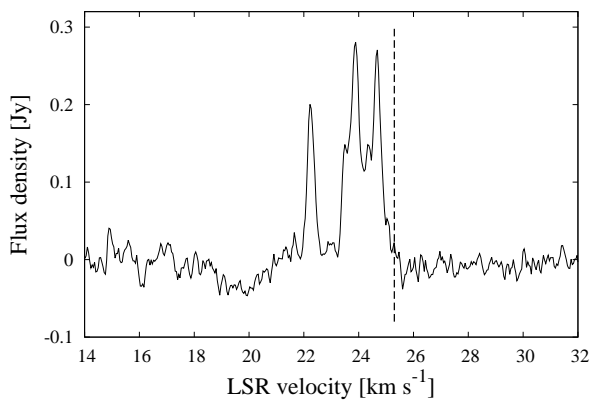} 
 \end{center}
\caption{
Spectrum in the quiescent phase defined in section~\ref{section3.4}. 
All the selected 561 scans of 5~min duration were used. 
The 3$\sigma$ detection limit was achieved to be 0.039~Jy. 
A dashed vertical line denotes the LSR velocity of the flaring component G 
(25.30~km~s$^{-1}$).
}\label{fig9}
\end{figure}

As a result, a spectrum in the quiescent phase was obtained from 
561 scans of 5~min duration, or $\sim$46.8~hrs (2,805~min) 
integration time in total (figure~\ref{fig9}), 
under an assumption that the emission in quiescent intervals was stable. 
A detection limit (3$\sigma$) was achieved to be 0.039~Jy. 
Five components whose flux densities were 0.15--0.3~Jy 
were detected in the $V_{\mathrm{lsr}}$ of 22--25~km~s$^{-1}$, 
roughly supposed as the components C--F. 
On the other hand, no emission was detected at the $V_{\mathrm{lsr}} =$ 
25.30~km~s$^{-1}$ of the flaring component~G (dashed line in figure~\ref{fig9}). 
We concluded that the ratio of the peak flux density in the flaring and quiescent phases 
was more than 180 when the flare showed the maximum peak flux density of $\sim$7~Jy.

\subsection{Comparison with other flaring sources}\label{section3.5}

We compare the flaring activities detected in the 6.7~GHz methanol maser 
of G~014.23 
to flares observed in other 6.7~GHz maser sources. 
First, in the point of view of 
the ratio of the peak flux density in the flare and quiescence, 
the flares at the component G in G~014.23 
can be compared to a spectral component 
of $V_{\mathrm{lsr}} =$ $-$5.88~km~s$^{-1}$ in G~351.42$+$00.64 
(NGC6334~F: \cite{goedhart04}) 
and those of $V_{\mathrm{lsr}} =$ $-$1.32 and $-$0.55~km~s$^{-1}$ 
in Cepheus~A \citep{szymczak14}. 
The spectral components in both sources showed 
ratios of the peak flux density in the flare of 170 and 240 
in comparison with the quiescent phase. 
It is also similar 
that only a part of spectral components show remarkable flux variability. 
The time scale of the flux rising, however, is different, 
i.e., approximately two and one months in G~351.42$+$00.64 and Cepheus~A 
respectively, while two days or less in G~014.23. 
Second, in the point of view of the time scale of the flux rising, 
the flares in G~014.23 can be compared to flaring activities 
of a spectral component 
of $V_{\mathrm{lsr}} =$ 59.6~km~s$^{-1}$ in G~033.64$-$00.22 
that showed the time scale of the flux rising of one day or less 
(\cite{fujisawa12}, \yearcite{fujisawa14b}). 
Ratios of the peak flux density in the flares of 
G~033.64$-$00.22 were, however, 
$\sim$7--27 in comparison with the quiescent phase 
and approximately one order of magnitude less than that in G~014.23. 
Finally, in the point of view of periodicity, 
the periodic flares in G~014.23 can be compared to 
periodic flaring activities detected in IRAS~22198$+$6336 
that showed the period of 34.6~days \citep{fujisawa14a}. 
Both the time scale of flux rising is comparable as short as 
two to three days or less. 
A ratio of the peak flux density in the flare of 
IRAS~22198$+$6336 was, however, 
more than 30 in comparison with the quiescent phase, 
i.e., one-fifth of the case in G~014.23, 
though it might be affected by poor detection limits (3$\sigma$) of $\sim$1.3~Jy 
in observations for IRAS~22198$+$6336. 
Furthermore, periodic flares occurred on all the spectral components 
with the time lag of two days in IRAS~22198$+$6336, 
while on only the single spectral component G in G~014.23. 

We concluded that the flaring activities detected in G~014.23 
are remarkable phenomena 
because of the large ratio of the peak flux density 
in comparison with the quiescent phase, 
the short time scale of the flux rising, 
the occurrence on just one spectral component, and the short periodicity.

\section{Discussions}\label{section4}

We discuss possible mechanisms to explain 
the intermittent periodic flares in the component G of G~014.23, 
including their 
ratio of the peak flux density more than 180 
in the flaring and quiescent phases, 
the time scale of the flux rising as short as two days, 
either the symmetric or asymmetric profile in each flare, 
and the period of 23.9~days. 

It was reported that interstellar scintillation caused flux variability 
on the time scale as short as minutes observed in hydroxyl masers 
\citep{clegg91}. 
This model, however, explains amplitude fluctuations with at most 20{\%} 
and is not suited to the flares in G~014.23. 
Furthermore, if the scintillation causes the flares, 
all the spectral components would show flux variability with the same pattern. 
Therefore, the flares were not caused by interstellar scintillation. 

The amplifying effect by overlapping two maser clouds 
along the line of sight \citep{deguchi89} could cause flaring activities 
and explain distinct ratio of flux densities 
with one magnitude or more brighter in comparison with quiescent phases. 
This effect was verified toward the outburst of the 22.2~GHz water maser 
in Orion~KL \citep{shimoikura05}. 
The period of 23.9~days, however, is too short 
unless the maser emission would directly originate from 
the circumstellar disk in Keplerian rotation. 
Assuming that the central mass is 3.3~\Mo \citep{povich10} and 
8~\Mo (typical one and the lower limit as a high-mass star), 
the period of 23.9~days is achieved at the radius of 0.2 and 0.3~au 
in the Keplerian disk, respectively. 
At these radii, the dust temperature $T_{\mathrm{dust}}$ should be higher than 1,000~K 
that is out of range for the 6.7~GHz methanol 
maser excitation 100~K~$\leq T_{\mathrm{dust}} \leq 300$~K \citep{cragg05}. 
Therefore, the overlapping model was not applicable to 
the periodic flares in G~014.23 with the period of 23.9~days. 

The magnetic reconnection around a central star 
can release the energy to heat the gas and dust localized at a flaring area 
and explain flaring activities 
with rising time scale of a few days or less, 
such as ones in G~033.64$-$00.22 (\cite{fujisawa12}, \yearcite{fujisawa14b}). 
The periodic flux variations persisting at least in 47 cycles 
as shown in the lower-panel of figure~\ref{fig3} 
and table~\ref{tab2}, however, 
can be hardly explained by the magnetic reconnection. 

The stellar pulsation of a high-mass YSO can produce 
the periodic flux variability through the change of dust temperature 
in the maser clouds \citep{inayoshi13}. 
All the spectral components should be affected in this pulsation model. 
The model of combining a constant and an independent pulsating YSOs 
is suggested to explain a periodic flux variation occurred in part of the components \citep{sanna15}. 
However, it is difficult to produce the intermittent pattern 
because the pulsation is driven by the $\kappa$ mechanism \citep{vanderwalt16}. 
Therefore, the periodic flares in G~014.23 
that occurred only on one component G showing the intermittent pattern 
cannot be explained by the pulsation model. 

The circumbinary disk model is also proposed to explain the periodic variability 
in some methanol sources through variations of the dust temperature 
\citep{araya10}. 
However, 
this model is not suited to the periodic flares 
in G~014.23 since it is difficult to cause periodic variations 
toward only one in seven spectral components 
as in the case of the pulsation model. 

The CWB model can produce the intermittent flux variability 
through the periodic change of the seed photon flux 
caused by periodically forming shocks 
at periastron due to collision of stellar winds (\cite{vanderwalt09}; \cite{vanderwalt11}). 
The ionizing radiation that originates from the shocks 
propagates through an H\emissiontype{II} region and changes the electron density 
resulting in periodic variability of the free-free emission. 
There has been a non-detection of an H\emissiontype{II} region 
in G~014.23 with typical rms noise level of $\sim$0.3~mJy~beam$^{-1}$ 
from the unbiased catalog for H\emissiontype{II} regions 
made through the CORNISH survey project 
(Co-ordinated Radio and Infrared Survey for High Mass Star Formation: 
\cite{hoare12}; \cite{purcell13}). 
The high-sensitivity survey (rms levels of $\sim$3--10~$\mu$Jy~beam$^{-1}$) 
using the Karl G. Jansky Very Large Array (JVLA), 
however, detected weak and compact radio continuum emissions toward 
58 high-mass star forming regions, most of which were non-detections 
in the CORNISH \citep{rosero16}. 
Although the YSO associated with G~014.23 whose evolutionary phase is classified 
into the stage 0/I may be too early to form H\emissiontype{II} regions, 
we need to conduct a high-sensitivity observation 
using the JVLA for verification whether an H\emissiontype{II} region exists 
in G~014.23. 

The dust temperature variation of the accretion disk 
around a protobinary caused by periodic rotation of hot and dense 
spiral shock wave in the disk central gap \citep{parfenov14} 
might be another candidate for explanation 
of the periodic flares in G~014.23, although some properties of this model 
might be improved in the suggestion by \citet{vanderwalt16}. 
This model can cause a flare toward only maser components 
localized along the line of sight between the central YSO 
and the maser-emitting region, 
which are illuminated by the hot component associated 
with the spiral shock as an extra heating source. 
The extra heat results in increment of column densities suitable for the maser excitation 
and amplifying a maser. 
The spiral shock is periodically rotated by the binary system 
and thus can cause periodic flares of the maser. 
It was noted by \citet{vanderwalt16} that the spiral shock model 
could not account for the intermittent pattern in periodic variations 
due to a tail of the shock. 
This issue could be resolved by replacing the extra heating source 
of the spiral shock with a shock formed by the CWB system 
within the disk central gap. 
Another issue noted by \citet{vanderwalt16} was 
that the luminosity of the spiral shock might be too low to play a role 
as the extra heating source. 
This issue was predicted from densities of gases in the gap region and 
of postshocked gases in the spiral shock at least five orders of magnitude 
lower than that used in the spiral shock model, which were estimated 
from a hydrogen density at the inner edge of the circumbinary disk 
assumed by \citet{parfenov14}. 
The same issue occurs even in the case of an improved model 
by replacing with the shock by the CWB \citep{vanderwalt16}. 
The issue of the low luminosity is unresolved, but this may be overcome 
under the unsaturated state (see section~\ref{section3.2} and figure~\ref{fig6}), 
causing strong flux variability by a bit of increment of column densities 
expected from the exponential relationship with the path length of the masing gas. 

To evaluate applicability of the shock model by the CWB within the disk central gap, 
we estimated the orbital semi-major axis of the binary to be 
0.26 and 0.34~au from the period of 23.9~days in the case of 
the mass of a central primary YSO 3.3~\Mo and typical high-mass star of 8~\Mo 
with the companion mass of 1~\Mo, respectively. 
The orbital semi-major axis never interferes with the primary star having the radius 
of 0.07 and 0.08~au estimated in the case of the mass of 3.3 and 8~\Mo, respectively, 
under the condition of the mass accretion rate 10$^{-4}$~\Mo~yr$^{-1}$ 
(see equation (13) of \cite{hosokawa09}). 
In the point of view of time scale of the flux rising, 
the dust in the maser emitting region in the disk is heated 
mostly by a diffuse emission from the ionized gas within the inner disk regions 
that in turn is re-processed stellar and shocked gas radiation. 
The diffuse emission produces the heating rate 
of order of 10$^{-13}$~erg~s$^{-1}$~cm$^{-3}$, 
resulting in the time scale of the order of a few hours for increasing 
$T_{\mathrm{dust}}$ by 10~K \citep{parfenov14}. 
The time scale for increasing $T_{\mathrm{dust}}$ by 100~K, therefore, 
is expected to be a few tens of hours, 
which is comparable to the time scale of the flux rising for the flaring 
component G, 
i.e., 21.3~hrs (1,278~min). 
While the profile of the flare in the period I of the intraday monitoring is 
symmetric with respect to its peak, 
those in the periods II and IV (and possibly in the period III) are asymmetric. 
These differences might be attributed to an inclination angle of the binary 
suggested in three-dimensional numerical models \citep{sytov09}. 
In the numerical model, the variation profile of the column density 
depends on the inclination angle of the binary, 
possibly shaping the profile of flares. 
Both symmetric and asymmetric profiles might happen 
if the inclination of the binary is close to the critical value for the inclination. 
To verify the inclination angle, 
we plan to obtain a spatial distribution of the maser emissions in G~014.23 
through a very-long-baseline-interferometry (VLBI) observation. 
For instance, if we obtain elliptical morphology, the inclination angle 
can be estimated under an assumption that all the masers are distributed 
on a concentric circle in a disk. 

We conclude that 
the models of 
the change in the flux of seed photons by the CWB system 
or the variation of the dust temperature by the CWB shock 
within the disk central gap, 
in which 
the orbital semi-major axes of the binary are 0.26--0.34~au, 
might explain all the characteristics of the intermittent periodic flare 
of the component G in G~014.23: 
the ratio of the peak flux density in the flare 
was more than 180 in comparison with 
the quiescent phase, 
the time scale of the flux rising was as short as two days or less, 
the flare was decayed with either the symmetric or asymmetric profile in each flare, 
and the period was short as 23.9~days.

\section{Summary}\label{section5}

We presented periodic and flaring flux variability of the 
6.7~GHz methanol maser emission detected in the spectral component G 
($V_{\mathrm{lsr}} =$ 25.30~km~s$^{-1}$) in G~014.23$-$00.50 
through a long-term and highly frequent monitoring 
using the Hitachi 32-m radio telescope. 
The flaring activities have regularly occurred 
with the period of 23.9$\pm 0.1$~days 
at least in 47 cycles from 5 January 2013 to 21 January 2016 
(MJD 56297--57408), corresponding to $\sim$1,100~days. 
The period of 23.9~days is the shortest one observed in the masers 
at around high-mass YSOs so far. 
The flaring component normally fell below 
the detection limit (3$\sigma$) of $\sim$0.9~Jy. 
In the flaring periods, the component rose above the detection limit 
with the ratio of the peak flux density more than 180 in comparison with 
the quiescent phase, showing the intermittent periodic variability. 
The time-scale of the flux rising was typically two days or shorter, 
and the flaring components were decayed with 
either symmetric or asymmetric profile, 
which were revealed through the intraday monitoring. 
These characteristics might be explained in the models of 
the change in the flux of seed photons by the CWB system 
or the variation of the dust temperature by the CWB shock within 
the disk central gap, 
in which the orbital semi-major axes of the binary were estimated to be 0.26--0.34~au.

\begin{ack}
The authors are grateful to Naoko Furukawa for substantial contributions 
to this monitoring project. 
Thanks are due to all the staff and students both at Ibaraki University 
and those graduated from the university, 
and the staff of the Mizusawa VLBI observatory and 
the members of the Japanese VLBI Network team, 
for the development and operation of the Hitachi 32-m radio telescope. 
The authors would also like to thank the anonymous referee 
for useful and constructive suggestions to improve the paper 
and Dr. Gemma Busquet for kindly 
sharing their calibrated SMA data in the fits format with us. 
This work was financially supported in part by a Grant-in-Aid for Scientific 
Research (KAKENHI) from Japan Society for the Promotion of Science 
(JSPS), No. 24340034. 
\end{ack}


\begin{thebibliography}{}

\bibitem[Araya et al.(2010)]{araya10} 
Araya, E.~D., Hofner, P., Goss, W.~M., Kurtz, S., Richards, A.~M.~S., 
Linz, H., Olmi, L., \& Sewi{\l}o, M.\ 2010, \apjl, 717, L133 

\bibitem[Breen et al.(2015)]{breen15} 
Breen, S.~L., et al.\ 2015, \mnras, 450, 4109 

\bibitem[Busquet et al.(2013)]{busquet13} 
Busquet, G., et al.\ 2013, \apjl, 764, L26 

\bibitem[Busquet et al.(2016)]{busquet16} 
Busquet, G., et al.\ 2016, \apj, 819, 139 

\bibitem[Caswell et al.(2010)]{caswell10} 
Caswell, J.~L., et al.\ 2010, \mnras, 404, 1029 

\bibitem[Caswell et al.(2011)]{caswell11} 
Caswell, J.~L., et al.\ 2011, \mnras, 417, 1964

\bibitem[Clegg \& Cordes(1991)]{clegg91} 
Clegg, A.~W., \& Cordes, J.~M.\ 1991, \apj, 374, 150 

\bibitem[Cragg et al.(2005)]{cragg05} 
Cragg, D.~M., Sobolev, A.~M., \& Godfrey, P.~D.\ 2005, \mnras, 360, 533 

\bibitem[Cyganowski et al.(2008)]{cyganowski08} 
Cyganowski, C.~J., et al.\ 2008, \aj, 136, 2391-2412 

\bibitem[Deguchi \& Watson(1989)]{deguchi89} 
Deguchi, S., \& Watson, W.~D.\ 1989, \apjl, 340, L17 

\bibitem[Elmegreen and Lada(1976)]{elmegreen76} 
Elmegreen, B.~G., \& Lada, C.~J.\ 1976, \aj, 81, 1089 

\bibitem[Fujisawa et al.(2012)]{fujisawa12} 
Fujisawa, K., et al.\ 2012, \pasj, 64,  

\bibitem[Fujisawa et al.(2014a)]{fujisawa14a} 
Fujisawa, K., et al.\ 2014a, \pasj, 66, 78 

\bibitem[Fujisawa et al.(2014b)]{fujisawa14b} 
Fujisawa, K., et al.\ 2014b, \pasj, 66, 109 

\bibitem[Goedhart et al.(2003)]{goedhart03} 
Goedhart, S., Gaylard, M.~J., \& van der Walt, D.~J.\ 2003, \mnras, 339, L33 

\bibitem[Goedhart et al.(2004)]{goedhart04} 
Goedhart, S., Gaylard, M.~J., \& van der Walt, D.~J.\ 2004, \mnras, 355, 553 

\bibitem[Goedhart et al.(2009)]{goedhart09} 
Goedhart, S., Langa, M.~C., Gaylard, M.~J., 
\& van der Walt, D.~J.\ 2009, \mnras, 398, 995 

\bibitem[Goedhart et al.(2014)]{goedhart14} 
Goedhart, S., Maswanganye, J.~P., Gaylard, M.~J., 
\& van der Walt, D.~J.\ 2014, \mnras, 437, 1808 

\bibitem[Goldreich \& Kwan(1974)]{goldreich74} 
Goldreich, P., \& Kwan, J.\ 1974, \apj, 190, 27 

\bibitem[Green et al.(2010)]{green10} 
Green, J.~A., et al.\ 2010, \mnras, 409, 913 

\bibitem[Green et al.(2012a)]{green12a} 
Green, J.~A., Caswell, J.~L., Voronkov, M.~A., 
\& McClure-Griffiths, N.~M.\ 2012a, \mnras, 425, 1504 

\bibitem[Green et al.(2012b)]{green12b} 
Green, J.~A., et al.\ 2012b, \mnras, 420, 3108 

\bibitem[Hoare et al.(2012)]{hoare12} 
Hoare, M.~G., et al.\ 2012, \pasp, 124, 939 

\bibitem[Hosokawa \& Omukai(2009)]{hosokawa09} 
Hosokawa, T., \& Omukai, K.\ 2009, \apj, 691, 823 

\bibitem[Inayoshi et al.(2013)]{inayoshi13} 
Inayoshi, K., Sugiyama, K., Hosokawa, T., Motogi, K., 
\& Tanaka, K.~E.~I.\ 2013, \apjl, 769, L20 

\bibitem[Kondo et al.(2008)]{kondo08}
Kondo, T., Koyama, Y., Ichikawa, R., Sekido, M., Kawai, E. 
\& Kimura, M. 2008, J. Geod. Soc. Jpn, 54, 233

\bibitem[Lomb(1976)]{lomb76} 
Lomb, N.~R.\ 1976, \apss, 39, 447 

\bibitem[Maswanganye et al.(2015)]{maswan15} 
Maswanganye, J.~P., Gaylard, M.~J., Goedhart, S., Walt, D.~J.~v.~d., 
\& Booth, R.~S.\ 2015, \mnras, 446, 2730 

\bibitem[Maswanganye et al.(2016)]{maswan16} 
Maswanganye, J.~P., van der Walt, D.~J., Goedhart, S., 
\& Gaylard, M.~J.\ 2016, \mnras, 456, 4335 

\bibitem[Ohashi et al.(2016)]{ohashi16} 
Ohashi, S., Sanhueza, P., Chen, H.-R.~V., Zhang, Q., Busquet, G., 
Nakamura, F., Palau, A., \& Tatematsu, K.\ 2016, \apj, 833, 209

\bibitem[Okoh et al.(2014)]{okoh14} 
Okoh, D., Esimbek, J., Zhou, J.~J., Tang, X.~D., Chukwude, A., 
Urama, J., \& Okeke, P.\ 2014, \apss, 350, 657 

\bibitem[Olmi et al.(2014)]{olmi14} 
Olmi, L., Araya, E.~D., Hofner, P., Molinari, S., Morales Ortiz, J., 
Moscadelli, L., \& Pestalozzi, M.\ 2014, \aap, 566, A18 

\bibitem[Parfenov \& Sobolev(2014)]{parfenov14} 
Parfenov, S.~Y., \& Sobolev, A.~M.\ 2014, \mnras, 444, 620 

\bibitem[Peretto and Fuller(2009)]{peretto09} 
Peretto, N., \& Fuller, G.~A.\ 2009, \aap, 505, 405 

\bibitem[Pestalozzi et al.(2005)]{pestalozzi05} 
Pestalozzi, M.~R., Minier, V., \& Booth, R.~S.\ 2005, \aap, 432, 737 

\bibitem[Povich \& Whitney(2010)]{povich10} 
Povich, M.~S., \& Whitney, B.~A.\ 2010, \apjl, 714, L285 

\bibitem[Purcell et al.(2013)]{purcell13} 
Purcell, C.~R., et al.\ 2013, \apjs, 205, 1 

\bibitem[Robitaille et al.(2006)]{robitaille06} 
Robitaille, T.~P., Whitney, B.~A., Indebetouw, R., 
Wood, K., \& Denzmore, P.\ 2006, \apjs, 167, 256 

\bibitem[Rosero et al.(2016)]{rosero16} 
Rosero, V., Hofner, P., Claussen, M., et al.\ 2016, \apjs, 227, 25

\bibitem[Sanna et al.(2015)]{sanna15} 
Sanna, A., et al.\ 2015, \apjl, 804, L2

\bibitem[Scargle(1982)]{scargle82} 
Scargle, J.~D.\ 1982, \apj, 263, 835 

\bibitem[Schlingman et al.(2011)]{schlingman11} 
Schlingman, W.~M., et al.\ 2011, \apjs, 195, 14 

\bibitem[Sewilo et al.(2004)]{sewilo04} 
Sewilo, M., Watson, C., Araya, E., Churchwell, E., 
Hofner, P., \& Kurtz, S.\ 2004, \apjs, 154, 553 

\bibitem[Shimoikura et al.(2005)]{shimoikura05} 
Shimoikura, T., Kobayashi, H., Omodaka, T., Diamond, P.~J., 
Matveyenko, L.~I., \& Fujisawa, K.\ 2005, \apj, 634, 459 

\bibitem[Shirley et al.(2013)]{shirley13} 
Shirley, Y.~L., et al.\ 2013, \apjs, 209, 2 

\bibitem[Sun et al.(2014)]{sun14} 
Sun, Y., et al.\ 2014, \aap, 563, A130 

\bibitem[Sytov et al.(2009)]{sytov09} 
Sytov, A.~Y., Bisikalo, D.~V., Kaigorodov, P.~V., \& Boyarchuk, A.~A.\ 2009, 
Astronomy Reports, 53, 428 

\bibitem[Szymczak et al.(2011)]{szymczak11} 
Szymczak, M., Wolak, P., Bartkiewicz, A., 
\& van Langevelde, H.~J.\ 2011, \aap, 531, L3 

\bibitem[Szymczak et al.(2014)]{szymczak14} 
Szymczak, M., Wolak, P., \& Bartkiewicz, A.\ 2014, \mnras, 439, 407 

\bibitem[Szymczak, Wolak, and Bartkiewicz(2015)]{szymczak15} 
Szymczak, M., Wolak, P., \& Bartkiewicz, A.\ 2015, \mnras, 448, 2284 

\bibitem[Szymczak et al.(2016)]{szymczak16} 
Szymczak, M., Olech, M., Wolak, P., Bartkiewicz, A., 
\& Gawro{\'n}ski, M.\ 2016, \mnras, 459, L56 

\bibitem[Ukita et al.(1981)]{ukita81} 
Ukita, N., Kaifu, N., Chikada, Y., Miyaji, T., \& Miyazawa, K.\ 1981, 
\pasj, 33, 341 

\bibitem[van der Walt et al.(2009)]{vanderwalt09} 
van der Walt, D.~J., Goedhart, S., \& Gaylard, M.~J.\ 2009, 
\mnras, 398, 961 

\bibitem[van der Walt(2011)]{vanderwalt11} 
van der Walt, D.~J.\ 2011, \aj, 141, 152 

\bibitem[van der Walt et al.(2016)]{vanderwalt16} 
van der Walt, D.~J., Maswanganye, J.~P., Etoka, S., 
Goedhart, S., \& van den Heever, S.~P.\ 2016, \aap, 588, A47 

\bibitem[Wang et al.(2006)]{wang06} 
Wang, Y., Zhang, Q., Rathborne, J.~M., Jackson, J., 
\& Wu, Y.\ 2006, \apjl, 651, L125 

\bibitem[Xu et al.(2009)]{xu09} 
Xu, Y., Voronkov, M.~A., Pandian, J.~D., Li, J.~J., Sobolev, A.~M., 
Brunthaler, A., Ritter, B., \& Menten, K.~M.\ 2009, \aap, 507, 1117 

\bibitem[Xu et al.(2011)]{xu11} 
Xu, Y., Moscadelli, L., Reid, M.~J., Menten, K.~M., Zhang, B., 
Zheng, X.~W., \& Brunthaler, A.\ 2011, \apj, 733, 25 

\bibitem[Yonekura et al.(2016)]{yonekura16} 
Yonekura, Y., et al.\ 2016, \pasj, 68, 74 

\end{thebibliography}
\end{document}